\documentstyle[12pt,epsf]{article}
\def\gsim{\mathrel{\rlap {\raise.5ex\hbox{$ > $}}
{\lower.5ex\hbox{$\sim$}}}}
\def\lsim{\mathrel{\rlap {\raise.5ex\hbox{$ < $}}
{\lower.5ex\hbox{$\sim$}}}}

\newcommand{\be}{\begin{equation}}
\newcommand{\ee}{\end{equation}}
\newcommand{\bea}{\begin{eqnarray}}

\newcommand{\eea}{\end{eqnarray}}
\newcommand{\nd}[1]{/\hspace{-0.6em} #1}

\def\gappeq{\mathrel{\rlap {\raise.5ex\hbox{$>$}}
{\lower.5ex\hbox{$\sim$}}}}
 
\def\lappeq{\mathrel{\rlap{\raise.5ex\hbox{$<$}}
{\lower.5ex\hbox{$\sim$}}}}
 
\begin{document}
 
\begin{titlepage}
\begin{flushright} 
gr-qc/9905048
\end{flushright} 

\begin{centering} 
{\large {\bf Search for Quantum Gravity}} \\ 
\vspace{.05in}

 {\bf John Ellis }$^{a}$, 
{\bf N.E. Mavromatos}$^{b,\diamond}$ 
and {\bf D.V. Nanopoulos}$^{c}$

\vspace{0.05in}
 
{\bf Summary} \\
\vspace{.1in}
\end{centering}
{\small 
A satisfactory theory of quantum gravity may necessitate 
a drastic modification of our perception of space-time,
by giving it a foamy structure
at distances comparable to the Planck length.
It is argued in this essay that the experimental 
detection of such structures may be a realistic possibility
in the foreseeable future.  
After a brief review of different theoretical
approaches to quantum gravity and the relationships
between them, we discuss various possible experimental tests of the
quantum nature of space-time. Observations of photons from distant
astrophysical sources such as Gamma-Ray Bursters and laboratory
experiments on neutral kaon decays may
be sensitive to quantum-gravitational effects 
if they are only minimally 
suppressed. Experimental limits from the Whipple Observatory and the
CPLEAR Collaboration are already probing close to the Planck scale, and
significant increases in sensitivity are feasible.}

\vspace{0.5in}
\begin{centering} 

{\it Awarded First Prize in the Gravity Research Foundation 
Essay Competition for 1999}

\end{centering} 

\vspace{0.50in}
\begin{flushleft}
$^{a}$ CERN, Theory Division, CH-1211 Geneva 23, Switzerland:
{\tt John.Ellis@cern.ch}\\
$^{b}$ University of Oxford, 
Department of Physics, Theoretical Physics,
1 Keble Road,
Oxford OX1 3NP, U.K.:
{\tt n.mavromatos1@physics.oxford.ac.uk} \\
$^{c}$ Center for Theoretical Physics,
Department of Physics, Texas A \& M University, College Station,
TX 77843-4242, U.S.A., \\
Astroparticle Physics Group, Houston Advanced Research Center (HARC),
The Mitchell Campus, The Woodlands, TX 77381, U.S.A., \\
Academy of Athens, Chair of Theoretical Physics, 
Division of Natural Sciences, 28 Panepistimiou Ave., 
Athens GR-10679, Greece: {\tt dimitri@soda.physics.tamu.edu} \\
$^{\diamond}$ P.P.A.R.C. Advanced Fellow.

\end{flushleft}

\end{titlepage} 

Almost a century has elapsed since Einstein proposed his
General Theory of Relativity, in which
the curvature of space encodes 
the classical gravitational field. Somewhat later, first
quantum mechanics and then quantum field theory were formulated.
All of these theories have been individually tested with
great accuracy.
However, a consistent quantum version of gravity still
eludes us, and it is often thought that quantum gravity
must lie beyond present experimental reach.

Attempts to quantize General Relativity may be
fitted into three major categories. 
One tackles the quantization 
of the geometry of space and time within 
the framework of local four-dimensional field theories 
and (non-trivial) extensions
thereof, using a {\it canonical formalism} such as
the loop-gravity approach~\cite{ashtekar},
in which the states of the 
theory are represented as functions of spin networks, leading
to a `polymer' structure of quantum space-time.

The second major category posits
a {\it foamy structure} of quantum space-time~\cite{hawk}.
in which Planck-size topological
fluctuations resembling black holes, with
microscopic event horizons,
appear spontaneously out of the vacuum and subsequently
evaporate back into it. 
The microscopic black-hole horizons
are viewed as providing a sort of `environment' that might induce
quantum decoherence of apparently isolated matter
systems~\cite{ehns,zurek}. These are described by density matrices
$\rho$ with `in' and `out' states that evolve in a manner
reminiscent of the quantum mechanics of 
open systems~\cite{ehns}:
\be
\partial _t \rho = i [\rho, H] + \nd{\delta H}\rho 
\label{ehnseq}
\ee
where $H$ is the Hamiltonian, and the 
matrix $\nd{\delta H}$, which has a non-commutator 
structure, represents collectively
quantum-gravity 
effects. In this picture, as in the canonical approach, Lorentz 
covariance may be lost in the splitting between 
the matter system and the quantum-gravitational `environment'.
Such a breaking of Lorentz covariance could be considered 
a property of the quantum-gravitational ground state, and therefore a
variety of spontaneous breaking.

The third category includes {\it string theory} and its 
non-perturbative D(irichlet)-brane extension~\cite{strings}.
The discovery of new non-local solitonic structures (membranes)
in string theory has
led to a new interpretation of the quantum space-time:
D-branes appear as space-time defects, which give 
rise to a `discrete'
cellular structure in the space-time manifold, in a spirit reminiscent
of the loop-gravity formalism. Multiple D-branes may overlap and
interact via the exchanges of open strings with ends attached
to the brane surface, yielding a non-commutative geometry
of space~\cite{witten,ms}. 

Further intriguing possible connections between these apparently disparate
approaches to quantum gravity have emerged recently. For example, there
are conceptual and possibly observational similarities 
between a `weave' state in the loop-gravity approach and one
formulation of space-time foam~\cite{ehns}. Moreover, the latter may be
reformulated in the D-brane approach~\cite{emn2}. This is because
the scattering of ordinary matter, in the presence 
of a microscopic `singular' fluctuation in space-time, 
requires a quantum treatment of the `recoil' of the corresponding
space-time defect. In string theory, one represents matter as 
closed string and the defect as a D-brane~\cite{strings}, 
whose recoil is not described simply 
by a conformal string background, but rather by a change in 
the background~\cite{emn2,kmw} over which the string propagates. 
The resulting string theory becomes `non-critical'~\cite{ddk},
flowing from one conformal background to another. 
This flow is a `non-equilibrium' process, which allows for the 
formation and evaporation of black holes in a string theory 
framework~\cite{emn2}, and a loss of coherence as argued
previously in the framework of space-time foam.  
This point of view is in agreement with the argument of~\cite{myers},
in the context of the D-brane approach to black holes~\cite{strings}, 
that pure quantum states cannot form black holes, 
implying that the formation
and evaporation of black
holes must be understood within the framework of quantum decoherence.  

The central feature of non-critical string is the
appearance of a Liouville field on the world sheet,
which we identify as a dynamical renormalization scale
that we can in turn identify as the
physical time~\cite{emn2,ms}. Quantum fluctuations in
the space-time background, that are represented by couplings on the
string world sheet, induce renormalization
via the Liouville field. The corresponding 
renormalization-group equation has precisely the form
(\ref{ehnseq}) postulated previously in the space-time-foam
approach. Moreover, the elevation of time to a
quantum variable leads to
non-trivial uncertainty relations between Liouville time and
and the collective space coordinates $Y^i$ of $D$ branes,
parallelling and extending the non-commutative geometry of~\cite{witten}. 

In the rest of this essay, we explore whether it may be
possible to test experimentally such ideas about the quantum-gravitational
structure of space-time. We are interested in signatures that are
characterized by deviations from conventional quantum mechanics
and quantum field theory, that would 
presumably be suppressed by some power
or exponent of the Planck Mass  $M_P \sim 10^{19}$ GeV. 
As we discuss below, several such effects may be at the edge
of observability if the suppression is just by a single power
of $M_P$. This might indeed be the case, since
the extra term $\nd{\delta H}$ in (\ref{ehnseq}) may have the 
generic magnitude ${\cal O} (E^2/M_P)$~\cite{US}. Similar estimates
have been made in the contexts of black holes and D-branes~\cite{OTHERUS,emn2},
and in the loop-gravity approach~\cite{pullin}.

We discuss first the possible effects of a quantum-gravitational environment 
on the propagation of a massless particle such as a photon.
The recoil of a massive space-time defect, modelled as a
D-brane, curves space-time, giving rise to 
a  gravitational field of the form~\cite{emn2}: 
\be 
    G_{ij} \sim \eta_{ij} + {\cal O}(\frac{E}{M_P}) 
\label{gravitrecoil}
\ee
where $E \ll M_P$ is the photon energy, and $\eta_{ij}$ is a 
flat Minkowski metric. The most important effect of such a distortion
of space-time is the appearance of an induced index of   
refraction: the effective (group) velocity $v$ of photons in the 
quantum-gravitational `medium' depends linearly on energy~\cite{emn3}
\be
    v = c\left(1 -{\cal O}(\frac{E}{M_P})\right)
\label{refractive}
\ee
where $c$ is the light velocity in empty space, 
and the minus sign reflects the fact that 
there is no superluminal propagation
in the D-brane recoil approach to stringy quantum
gravity~\cite{ms,emn3,NEXTUS}. Such an index of refraction
has an energy dependence that is quite distinct from that in a 
conventional electromagnetic plasma, which decreases with increasing 
energy.

An analogous effect may arise in the loop approach 
to quantum gravity~\cite{ashtekar}, if the gravitational
degrees of freedom are in a ``weave'' state $|\Delta >$:
\be
<\Delta | G_{ab} |\Delta > = \eta_{ab} + 
{\cal O}\left(\frac{1}{M_P \Delta}\right) 
\label{weave}
\ee
where $\Delta$ is a
characteristic
length scale of the system~\cite{pullin}.
Maxwell's equations for
the propagation of ordinary photons are modified in the presence
of such a weave state (\ref{weave}), leading to
a modified index of refraction of the form (\ref{refractive}).
Novelties in the loop-gravity case (\ref{weave}) include the
possibility of superluminal propagation and a dependence
on the helicity of the photon state, which could
lead to characteristic birefringence effects. 

Finally, we note that photons with the same
energy (frequency) might travel at different velocities,
as is suggested by higher-order studies in stringy quantum
gravity~\cite{NEXTUS}. This would provide a second possible source of
dispersion
in a wave packet, beyond that associated with differing frequencies.

It is exciting that the existence 
of a non-trivial index of refraction or other possible
modification in the propagation of photons,
due to their interaction with a quantum-gravitational medium,
might be testable in the near future, if a 
suppression $E/M_{QG}$ is valid, with $M_{QG} \sim M_P$. The figure of
merit for such tests
is $(L \times E)/ \Delta t$, where $L$ is the distance of a
source of photons of energy $E$ which exhibits structure on
a time scale of order $\Delta t$. As was pointed out in~\cite{aemns},
gamma-ray bursters (GRBs) may have particularly large figures of merit,
as some exhibit microstructures around a millisecond, they may emit
$\gamma$ rays in the GeV or even TeV range, and many are now
known to be located at cosmological distances. It was estimated
in~\cite{aemns} that GRB observations might already be sensitive to a
quantum-gravity scale $M_{QG} ~\sim 10^{16}$~GeV, and 
suggested that the HEGRA and
Whipple air Cerenkov telescopes might be able to improve this
sensitivity. The Whipple group has now applied this idea to observations
of the active galaxy Markarian 421, establishing a lower
limit $M_{QG} > 4 \times 10^{16}$~GeV~\cite{biller}. A
possible HEGRA observation of high-energy $\gamma$ rays from GRB 920925c
might be sensitive to $M_{QG} ~\sim 10^{19}$~GeV~\cite{aemns2}, and
sensitive future tests could be made with the space
experiments AMS and GLAST.

Laboratory experiments with elementary particles
may also be used to probe the possible
quantum nature of space-time, as parametrized
by the modified time-evolution
equation (\ref{ehnseq}), for example in the neutral
kaon system~\cite{ehns,emn2,elmn}. 
Data from the CPLEAR collaboration have been used~\cite{emncplear}
to set upper limits on the possible decohering effects of the
quantum-gravitational environment at the level of 
$1 / ( 10^{17}~{\rm to}~10^{20})$~GeV, and there are
prospects for improving these limits in
future experiments on neutral kaons and mesons containing
bottom quarks. It has also
been suggested that interesting limits might be obtainable from
experiments on
neutrino oscillations~\cite{chinese}.

Finally, we point out the possibility that the non-commutative 
structure of space-time induced by multiple D-branes~\cite{ms}, as well as
modified uncertainty relations, 
might be detectable in atom interferometers~\cite{strunz}.
Based on the description of topological defects in 
space-time as D-branes~\cite{emn2,ms}, and the
non-trivial connection between D-particle recoil and diffusion 
in open systems~\cite{emn2}, it seems that
the non-commutativity of space-time
might indeed be testable in experiments of the type
discussed in~\cite{strunz}.
 
The above examples indicate that experimental tests of some ideas about
quantum gravity might not be so difficult as is often thought.
We have sketched in this essay an embryonic experimental strategy capable
of putting 
stringent bounds on quantum-gravitational effects, at least in certain
approaches.
The challenge for theorists now is to explore further  
the existing models, and to construct new ones that could 
provide a more complete guide to our experimental colleagues. 
The challenge for experimentalists is to prove these ideas wrong,
which may not be too difficult. The beginning of the next
millennium may already provide exciting opportunities to
seek quantum gravity.

\end{document}